\documentclass[%
 twocolumn,
 amsmath,amssymb,
 aps,
pra,
]{revtex4-2}

\usepackage{graphicx}
\usepackage{dcolumn}
\usepackage{bm}
\usepackage{ulem}  

\begin{document}

\preprint{arXiv:2201.09404}

\title{Mirror Coating Thermal Noise Mitigation Using Multi-Spatial Mode Cavity Readout}

\author{Andrew Wade}
 \email{andrew.wade@anu.edu.au}
\author{Kirk McKenzie}%
 \affiliation{%
The Centre for Gravitational Astrophysics, Research School of Physics, The Australian National University, Canberra ACT 2601, Australia
 }%

\date{\today}

\begin{abstract}
    We present an approach to mitigating coating thermal noise in optical cavities by using multiple TEM spatial modes to readout and stabilize laser frequency. With optimal weightings we synthesis a wider sampling of mirror surfaces, improving the averaging of Brownian thermal fluctuations. We show a thermal noise improvement factors of 1.57, comparable to a MESA beam of a nominal 12 m prototype cavity, and a factor 1.61 improvement over the $\textrm{TEM}_{00}$ using three modes in a 0.1 m cavity for a practical laboratory experiment: equivalent to cooling mirrors to 120 K from room temperature.
\end{abstract}

\maketitle


Stabilized lasers, referenced to ultra-stable optical cavities, are a mainstay of precision laser measurements in applications as diverse as narrow line width spectroscopy \cite{cygan_cavity_2013, ludlow_compact_2007}, atomic clocks \cite{jiang_making_2011, udem_optical_2002}, optical communication \cite{al-taiy_ultra-narrow_2014}, inter-spacecraft laser ranging \cite{abich-orbit_2019}, gravitational wave astronomy \cite{LIGOCollaboration2009, european_space_agency_laser_nodate} and tests of fundamental physics such as Lorentz Invariance \cite{herrmann_rotating_2009}. At low frequencies (below 100Hz) fundamental thermal noise directly limits the ultimately sensitivity of interferometic measurements and bounds the ultimate precision and accuracy of atomic frequency standards by limiting the interaction time in spectroscopic measurements \cite{quessada_dick_2003}. Addressing thermal noise is therefore essential for progress in a range of measurement and time keeping standards.

 Brownian coating thermal noise, arising from thermally driven fluctuations of mechanically lossy reflective coatings, represents the ultimate limit to laser stability at frequencies lower than 10 Hz \cite{NotcuttM2006}. This noise can be understood as random motion of the mirror surface that imparts a non-zero phase shift on incident laser light. The amplitude spectral density of this effect scales as $\sqrt{G_\nu(f)}\propto \nu_0\cdot 1/L_\textrm{cav}\cdot\sqrt{k_BT}\cdot\sqrt{\phi}\cdot 1/w_0\cdot\sqrt{1/f}$, where ${G_\nu(f)}$ is the power spectral density of the frequency noise, $\nu_0$ is the laser absolute frequency, $L_\textrm{cav}$ is the cavity length, $k_B$ is the Boltzmann constant, $T$ is absolute temperature, $\phi$ is the loss angle (the ratio of the imaginary to the real part of the elastic constant), $w_0$ is the effective beam radius on the mirrors, and $f$ is the Fourier frequency\cite{NumataK2004}. All materials exhibit multiple loss angles, each with temperature and frequency dependence, that characterize the thermal noise \cite{Hong2013}. Steady progress has been made in reducing mechanical loss angles ($\phi$) through coating materials and methods. Amorphous ion beam spluttered coatings have seen a factor of two improvement in their expected thermal noise amplitude \cite{vajente_low_2021} and crystalline coatings have shown three-fold reductions \cite{cole_tenfold_2013}. Direct reductions in thermal noise are also achieved with cryogenic cooling \cite{robinson_crystalline_2019}, though this not realizable in all applications due to limited material choices for some optical wavelengths, materials with loss angles with inverse temperature relationships, and, in demanding applications such as space missions where weight and complexity are a concern. Lengthening of cavities and shorter wavelengths also offer obvious improvement but are challenged in their ultimate enhancement by cavity mechanical stability as well as practical limitations in short wavelength laser sources and materials. The finally remaining approach, optimisation of the effective beam size on mirrors is a complementary strategy that offers a multiplying factor of improvement in addition to the aforementioned approaches and is the focus of this work.

\begin{figure}
    \centering
    \includegraphics[width=0.45\textwidth]{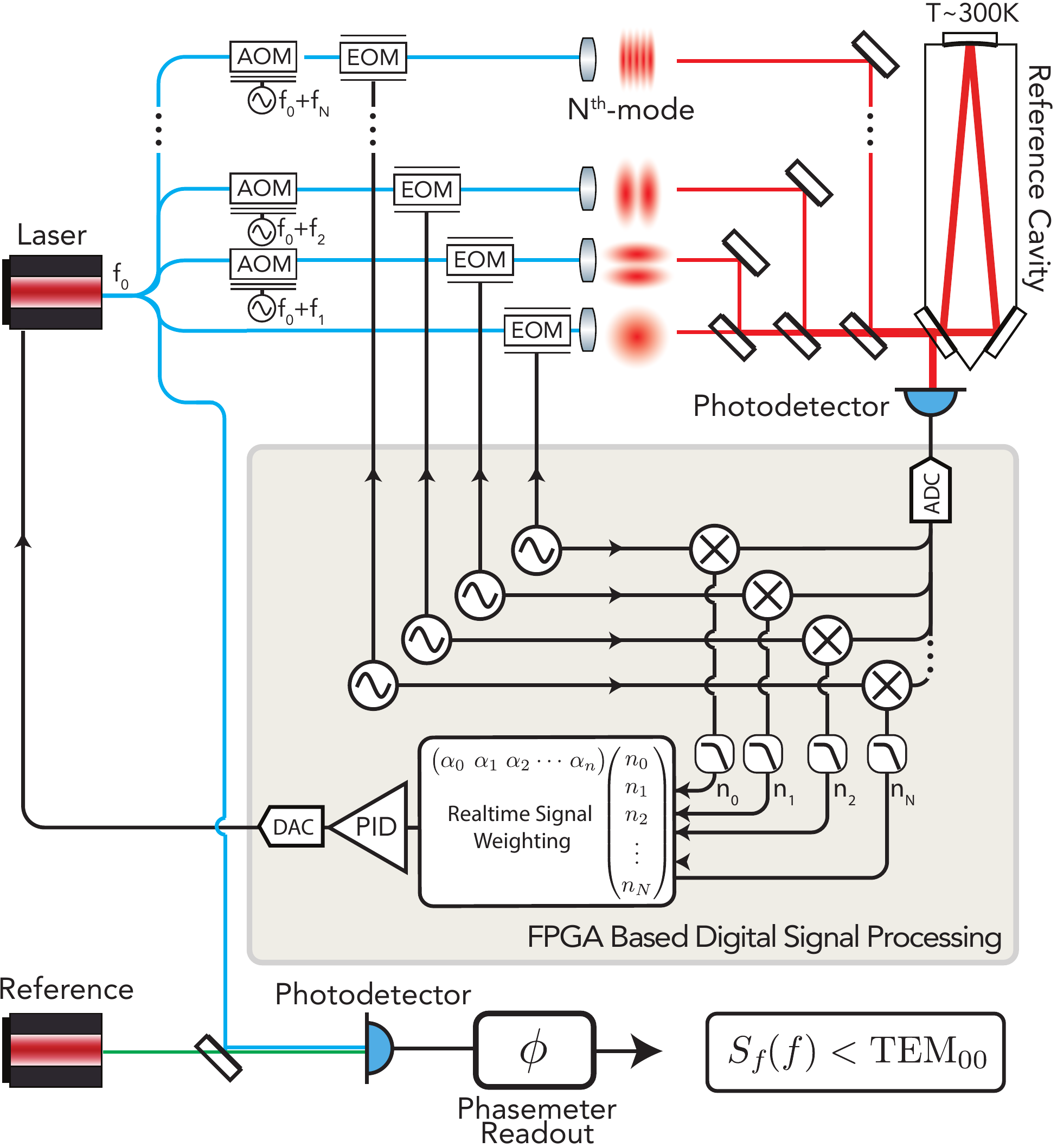}
    \caption{The proposed multi-mode readout scheme. Light from a laser is frequency shifted to match the higher order mode resonances of a reference cavity. Each frequency component is phase modulated with a unique RF frequency and injected.  Readout is made on a single reflection photo-detector and a Pound-Drever-Hall error signal is extracted for each mode in the cavity.  The resulting signals are combined with a weighting that optimizes canceling of thermal noise sampled by each unique mode profile.  The synthesized error signal spreads the effective sampled area of the mirrors.  The combined error signal is feed back to the laser, via a Proportional–Integral–Derivative (PID) controller, tying laser frequency to that of the optical resonator. Comparison of frequency noise performance is made with a beat note detection with phasemeter against a reference laser of identical or better performance.}
    \label{fig:HOMCavityConcept}
\end{figure}

We proposed a multi-mode cavity sensing technique that approaches the `MESA beam' thermal noise limit but avoids the currently insurmountable manufacturing requirements needed for such a system.  The MESA `flat-top' beam was proposed by O’Shaughnessy and Thorne as the ideal trade off point between a broadened intensity distribution and clipping optical power losses at the edge of mirrors\cite{oshaughnessy_topics_2003}. Though a functional demonstration has been reported \cite{tarallo_generation_2007}, manufacturing challenges in mirror profiles have stymied progress\cite{MillerPhD2010} such that no thermal noise improvement has been realized. However, approximation of such a flattened beam can be formed with spherical mirrors by the superposition of many basis modes of either the Hermite-Gauss or Laguerre-Gauss basis set. Shown conceptually in Fig.~\ref{fig:HOMCavityConcept}, a standard optical reference cavity is driven by several optical frequencies that are resonant for each unique higher order spatial mode, each of which is independently modulated and readout using the Pound-Drever-Hall technique \cite{drever_laser_1983}.  The resulting error signals are combined with optimal weightings to extract a composite signal with ideal combination to sample the largest possible effective mirror area, thereby minimizing coating thermal noise.  In contrast to other beam spreading strategies --- such as top-hat MESA modes \cite{oshaughnessy_topics_2003, vinet_mirror_2005}, edge stability cavities \cite{amairi_reducing_2013}, or, single higher order modes \cite{zeng_thermal-noise-limited_2018,sorazu_experimental_2013,chelkowski_prospects_2009,tao_higher-order_2020} --- this method allows for the use of conventional spherically polished mirrors with a greater robustness to manufacturing defects and alignment errors. This approach makes use of standard RF locking techniques but lends itself well to modern digital signal processing approaches where quantity of signals and flexible recombination may be easily scaled. 

This article demonstrates significant stabilization improvements through this multi-mode combination approach and defines a experimentally achievable system using a standard laser stabilization cavity and off-the-shelf components.  First we introduce a numerical simulation to validate the concept: initially with an arbitrary number of higher order modes (HOMs), showing an example improvement factor of 2.46 using the first 25 Hermite-Gaussian modes. Second, we focus on a direct comparison with a MESA beam generated in long aspect ratio cavities (similar to \cite{tarallo_generation_2007}) and show this technique delivers a similar thermal noise performance to a MESA beam, achieving 1.57 using the first 25 Hermite-Gauss modes. This is the regime most often encountered in interferometric gravitational wave detectors, for example Advanced LIGO\cite{LIGOCollaboration2009}, where arm cavities are of order tens to thousands of meters and beams fill the full mirror aperture and are limited by beam clipping.  Finally, we investigate a realisable experiment setup that shows a thermal noise improvement of 1.61 in a 10 cm cavity, using just three Hermite-Gaussian Modes --- the $\textrm{HG}_{00}$, $\textrm{HG}_{02}$, and $\textrm{HG}_{20}$ modes --- in what is an achievable experimental system. 

\textit{\textbf{Model---}} We construct the coating thermal noise model as follows: we break the mirror surfaces into $n\times m$ pixels, each of which is encoded with white uncorrelated noise\footnote{Coating Brownian thermal noise can be computed directly using the fluctuation dissipation theorem \cite{SaulsonP1990} and Levin's direct approach \cite{LevinY1998}.  There, the thermal Brownian motion associated with a mode profile of a particular optical mode shape is related directly to the loss of the of the system driven with a force profile matching the beam shape.  However, aspect ratio of beam size to coating thickness means that the strain energy calculation of any applied optical mode shape reduces to a surface integral \cite{vinet_thermal_2010}.  To first order localized strain deviations are unconnected to adjacent area domains. This characteristic means that it can be directly modeled as a array of statistically independent white noise elements across the surface of a mirror.  Though this approach is computationally intensive it is conceptually straightforward and sufficient to capture the essential scaling of Brownian coating noise with optical beam intensity distribution. This model is expected to break down as the pixels pitch and nodal features of incident HOM approach a length scale comparable to the coating thickness, at which point the surface integral approximation breaks down.}.  Each pixel is sampled at rate $f_s$ with time steps denoted by index $k$.  

The net fluctuation imparted onto an incident beam is computed by weighting the noise array by the intensity profile of the incident laser light. When sampling the cavity system with multiple HOMs the same noise mesh tensor, $N_{n, m}(k)$,  is used for each mode's intensity profile and the resulting composite readout signal $\Delta\tilde{\mathcal{L}}(k)$ is formed:
\begin{equation}\label{eq:CompoundControlSigCTN}
    \Delta\tilde{\mathcal{L}}(k) = \sum_{n, m, i}\alpha_i F_{n, m, i} N_{n,m}(k)
\end{equation}
where $\alpha_i$ is the weighing for the $i^\textrm{th}$ HOM probe read out from the cavity and $F_{n, m, i}$ is the normalized intensity of the $i^\textrm{th}$ HOM cavity mode. From this time series,  $\Delta\tilde{\mathcal{L}}(k)$, the data may be reshaped to have the characteristic $1/\sqrt{f}$ noise shape and the spectral densities may be computed for comparisons between modes and their different weighted re-combinations. Length variations, $\Delta \mathcal{L}$, are translated into laser frequency equivalent noise, $\Delta \nu$, by the usual relation, $\Delta \nu/\nu_0 = \Delta \mathcal{L}/L_\textrm{cav}$.

To calibrate the pixel-wise noise, the noise density is normalized against a sample Gaussian beam for a given pixel size. Key results presented in this article are ratios of improvement, so the ultimate choice of reference Brownian coating thermal noise level can be chosen arbitrarily. Here we use Numata et al.'s estimate \cite{NumataK2004} as a reference.  To verify the model we computed the noise at 1 Hz for a sweep of fundamental Gaussian mode beam sizes: this showed a characteristic $1/w_0$ scaling for beam radius $w_0$ \cite{nakagawa_thermal_2002}. Furthermore, we computed the Hermite-Gauss spatial mode intensity profiles for the first 36 modes to verify proper noise scaling. We find the higher order mode's thermal noise agreed with the direct Levin approach presented by Vinet (see Table II in \cite{vinet_thermal_2010}) to within the expected statistical variations of our Monty Carlo approach (less than 1.0\%).

\section{Case 1: Short Reference Cavity With No Beam Clipping Loss}
In this section we show results for short aspect ratio cavities with arbitrary numbers of modes.  For cavities length or order 0.1 m, the typical choices of beam sizes are dictated by geometric (cavity stability) factors, rather than considerations of power loss due to beam clipping at mirror edges.  When optimizing the combination of PDH error signal weightings, clipping losses are negligible in all but the most extreme higher order mode cases.  This scenario represents the best possible thermal performance for the case of unconstrained resources in the optical injection and readout signal processing.

To compare and optimize the relative weightings of control signals, the 1 Hz equivalent thermal noise was computed as a function of a weighting vector $\alpha_k$. The total RMS of the thermal noise is
$\Delta\tilde{\mathcal{L}}_\textrm{RMS} = \sqrt{1/Nf_s\sum_k\Delta\tilde{\mathcal{L}}(k)^2}$, where N is number of sample points, $f_s$ is the sample rate and $\Delta\tilde{\mathcal{L}}(k)$ is given in Eqn \ref{eq:CompoundControlSigCTN}. A noise tensor, $N_{n,m}(k)$, was generated and a standard non-linear solver (MATLAB's \texttt{fmincon}, interior-point algorithm) was used to discover the weighted combination of higher order modes control signals minimizing thermal noise.  Here we require that $\sum\alpha_i = 1$ (L1-norm) so that in the closed-loop feedback configuration the computed compound control signal remains properly normalized.
\begin{figure*}
    \centering
    \includegraphics[width=1.0\textwidth]{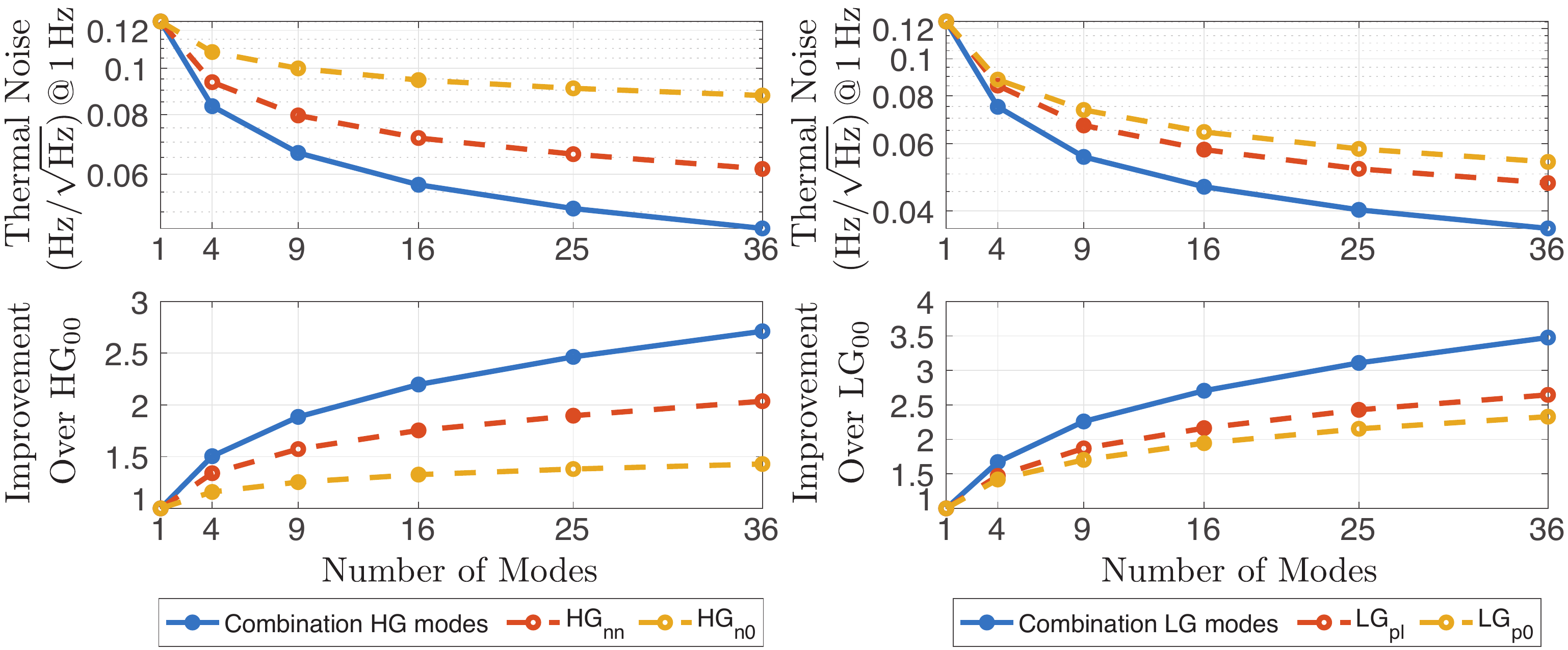}
    \caption{Thermal noise (upper plots) and thermal noise fractional improvement (lower plots) for combinations of Hermite-Gauss (left) and Laguerre-Gaussian (right) modes as compared to the fundamental mode. Cavity parameters: 0.1 m length with a 320 $\mu$m beam waist.  Thermal noise of single modes of highest order available are also plotted for comparison ($\textrm{HG}_{n0}$ and $\textrm{LG}_{p0}$). Improvement is unbounded in the absence of a clipping criteria and unlimited injection and readout resources.}
    \label{fig:TNImprovementNoClipSteppingNumModes}
\end{figure*}

\begin{figure}
    \centering
    \includegraphics[width=0.5\textwidth]{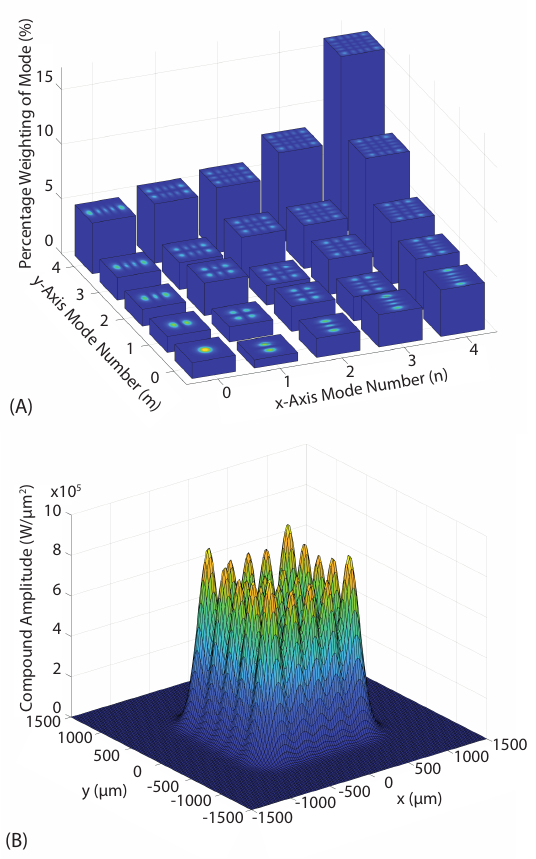}
    \caption{Example Hermite-Gaussian modes weightings for the case of short aspect ratio cavities in the non-clipping limited case. Optimally combined modes flatten the central region of the beam creating steeper cutoff at the edges. Here the total number of modes is 25, that is, limited to a maximum mode order number of four in both directions. The best weighting places most of the emphasis on the highest order modes available but lower order modes still make contributions to effective sampling of the mirror. Table \ref{tab:HOMNoiseRefTEM00} gives the values for this example mode combination.} 
    \label{fig:ModeWeightingInfographic}
\end{figure}

The solid lines in the top panels of Figure \ref{fig:TNImprovementNoClipSteppingNumModes} show thermal noise calculated using the multiple higher order mode sensing scheme for both the Hermite-Gaussian (left) and Laguerre-Gaussian mode (right) basis, as a function of available higher order modes. The solid line in the lower panels shows the improvement of the multiple mode sensing compared to the fundamental mode. The dashed traces show relative performance of using the highest order mode available for comparison.  Here, the waist of the resonant modes was set to 320 $\mu$m for a 0.1 m long cavity, ensuring a workable stability with g-factor of 0.81. Weightings of an example family of the first 25 Hermite-Gauss modes are shown in Figure \ref{fig:ModeWeightingInfographic}A (and Table \ref{tab:HOMNoiseRefTEM00}) with an effective sampled mode profile in Figure  \ref{fig:ModeWeightingInfographic}B. In the case of unconstrained experimental resources, that is, unlimited available modes, an optimal combination weights the outermost higher order modes the most. However, the interior node and anti-node structure of the higher order modes means that some of the lowest order modes help to flatten the weighting profile of the total effective sampling of the mirror. The improvement is $1.88\pm0.01$ for 9 modes and $2.46\pm0.01$ for 25 modes.
\begin{table}[]
    \centering
    \begin{tabular}{c|c|c|c|c|c}
    mode number n,m & 0 & 1 & 2 & 3 & 4 \\ \hline
    0 & 0.0147 & 0.0075 & 0.0175 & 0.0296 & 0.0427 \\
    1 & 0.0145 & 0.0137 & 0.0227 & 0.0219 & 0.0469 \\
    2 & 0.0156 & 0.0298 & 0.0180 & 0.0322 & 0.0563 \\
    3 & 0.0207 & 0.0244 & 0.0381 & 0.0394 & 0.0910 \\
    4 & 0.0466 & 0.0544 & 0.0594 & 0.0819 & 0.1604 \\
    \end{tabular}
    \caption{Hermite-Gaussian mode weightings for optimal thermal noise reduction using a combination of modes up to order 4 in the x (n) and y (m) directions for the non-beam clipping limited case in a short cavity. Weightings are approximately symmetric across the diagonal to within the statistical variance of the Monty Carlo model.}
    \label{tab:HOMNoiseRefTEM00}
\end{table}

\section{Case 2: Long aspect ratio cavities with beam clipping}
\label{sec:ClipLimitedCavities}
For optical resonators with an elongated aspect ratios, such as those present in interferometric gravitational wave detectors like the Advanced LIGO, beams propagate sufficient distance to fill the entire aperture of mirrors. The ideal beam size choice is constrained between the need to maximize beam profiles and that of limiting the intra-cavity loss due to clipping. Loss of this kind is undesirable as it leads to unwanted scattered light and lowers the effective finesse of the cavity. Constraining the optimization in this case is the requirement of thermal noise minimization, normalization of the control signal ($\sum \alpha_i = 1$) and a bound on allowable losses, that is, $\sum_i \alpha_i\varepsilon_i(\omega_0, R) \le\varepsilon_\textrm{Total}$. Here $\varepsilon_i(\omega_0, R)$ is the loss of the $i^\textrm{th}$ HOM injected for a given cavity waist of $\omega_0$ and mirror aperture radius $R$. $\varepsilon_\textrm{Total}$ is the total acceptable loss and places a bound on the allowable solutions.

\begin{figure}
    \centering
    \includegraphics[width=0.45\textwidth]{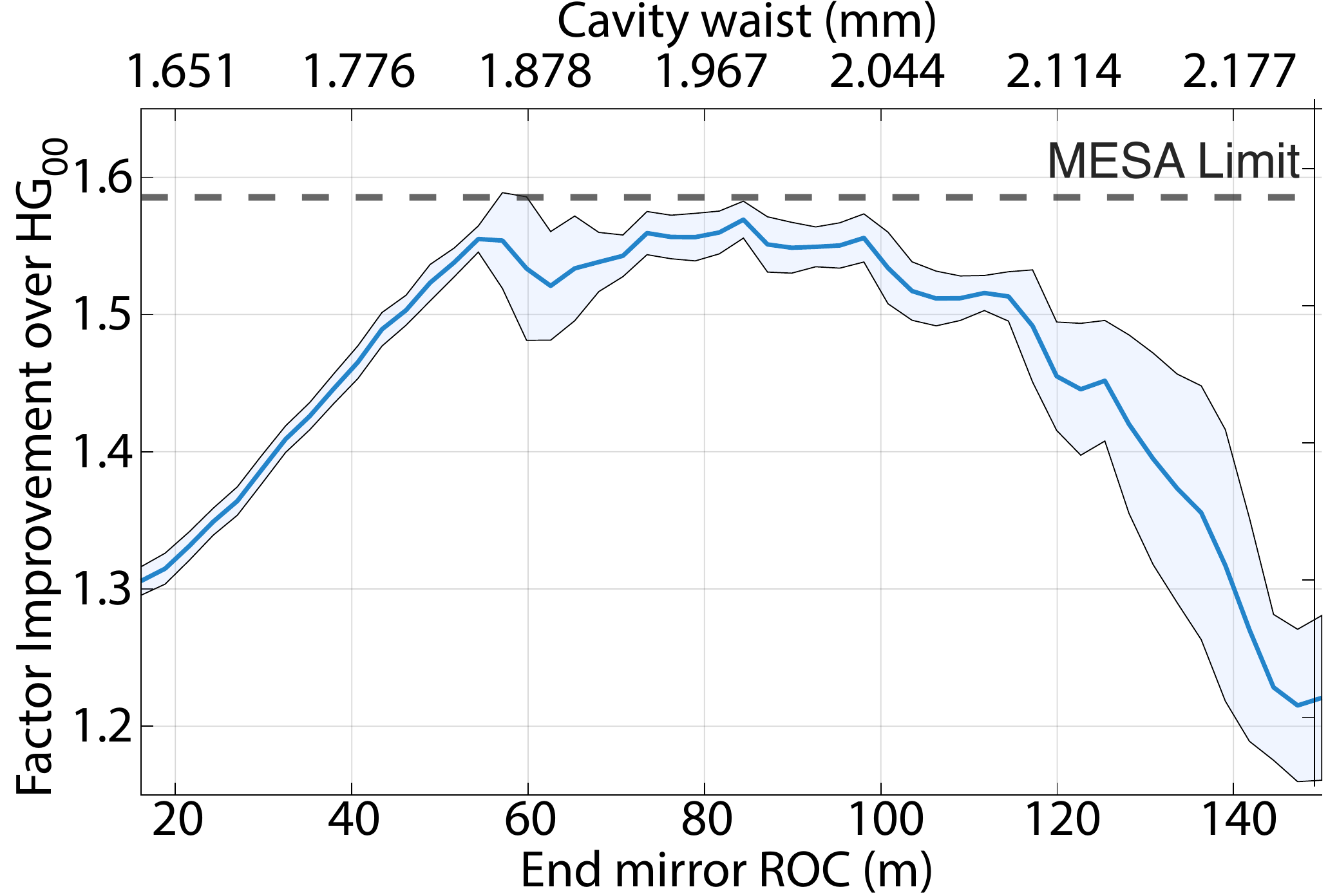}
    \caption{Ratio of improvement for an optimal combination of the first 25 Hermite-Gaussian modes over the clipping limited $\textrm{HG}_{00}$ fundamental mode as the beam waist is swept. The dark blue line gives mean value and the shaded region give the standard deviation.  Maximum improvement to thermal noise peaks at the MESA beam limit of 1.58. Here we use a 12 m cavity and compare against a MESA type beam (D=$4\sqrt{L_\textrm{cav}\lambda/2\pi}$), each clipping at the same 11.2 mm clipping aperture. The sweep is performed by stepping the end mirrors' radius of curvature and recomputing the thermal noise and optimal combination at each point. Combined losses are bound to less than 1 ppm.}
    \label{fig:SweepRComapairH00ExpandedToClip}
\end{figure}

With these constraints the optimization is widened to include a search over not only the weightings of the HOM control signals but also the choice of cavity geometry that determines the common beam waist between modes.  For comparison with the performance of MESA beams we choose a nominal cavity length of 12 m to match the best effort to date in producing top hat spread beams in optical cavities using modified mirrors\cite{tarallo_generation_2007}. Correspondingly we fix the mirror aperture to a radius of 11.2 mm and total effective loss tolerance to 1 ppm. As a figure of merit, thermal noise was computed at the 1 Hz point of the spectral density from a randomly generated noise tensor and the optimal weightings were computed as a function of end mirror Radius of Curvature (ROC): effectively sweeping the beam waist ($\omega_0$) to discover the best combination of beam focusing vs multi-mode combination.  Figure \ref{fig:SweepRComapairH00ExpandedToClip} shows the fractional improvement of thermal noise for the multi-mode readout over the clipping limited fundamental Gaussian beam. These results show that for a combination of the first 25 modes in the Hermite-Gauss mode basis the fractional improvement of $1.57\pm 0.01$ was achieved over the clipping limited fundamental Gaussian beam at ROC 84.4 m. For a MESA beam of disk radius, D, equal to four times the minimal Gaussian beam ($\sqrt{L_\textrm{cav}\lambda}/2\pi$, where $\lambda$ is wavelength), thermal noise improvement was a factor of $1.58\pm 0.01$, calculated from equation 2.4 in \cite{MillerPhD2010} and our Monty-Carlo model. The peak improvement is therefore comparable to a top-hat beam under equivalent constraints. The standard deviation in this fractional improvement verses radius of curvature sweep is of order $\pm$1\%, defined by the sampling resolution in the Monty-Carlo model and the optimizer's variance in weighting values between sample runs.

\section{Case 3: Practical Table-Top Experimental Mode Combinations}
For a practical first implementation of a multi-mode readout technique it is desirable to restrict the number of modes.  Limiting the total injected modes to three allows a manageable input optic chain but with the benefit of the largest marginal improved per mode added.  A combination of fundamental $HG_{00}$ and two higher order modes,  $HG_{02}$ and $HG_{20}$, was made in simulation. There are several experimental examples of order-two modes in use in cavity experiments \cite{gras_direct_2018, zeng_thermal-noise-limited_2018}. A choice of even ordered modes is ideal as their symmetry makes them first order insensitive to angular error and input pointing jitter.  They are also of a low enough order to be able to achieve reasonable levels of power coupling with no specialist optics, with up to 27\% overlap with fundamental Gaussian fields \cite{zeng_thermal-noise-limited_2018}.  \\

\begin{figure}
    \centering
    \includegraphics[width=0.48\textwidth]{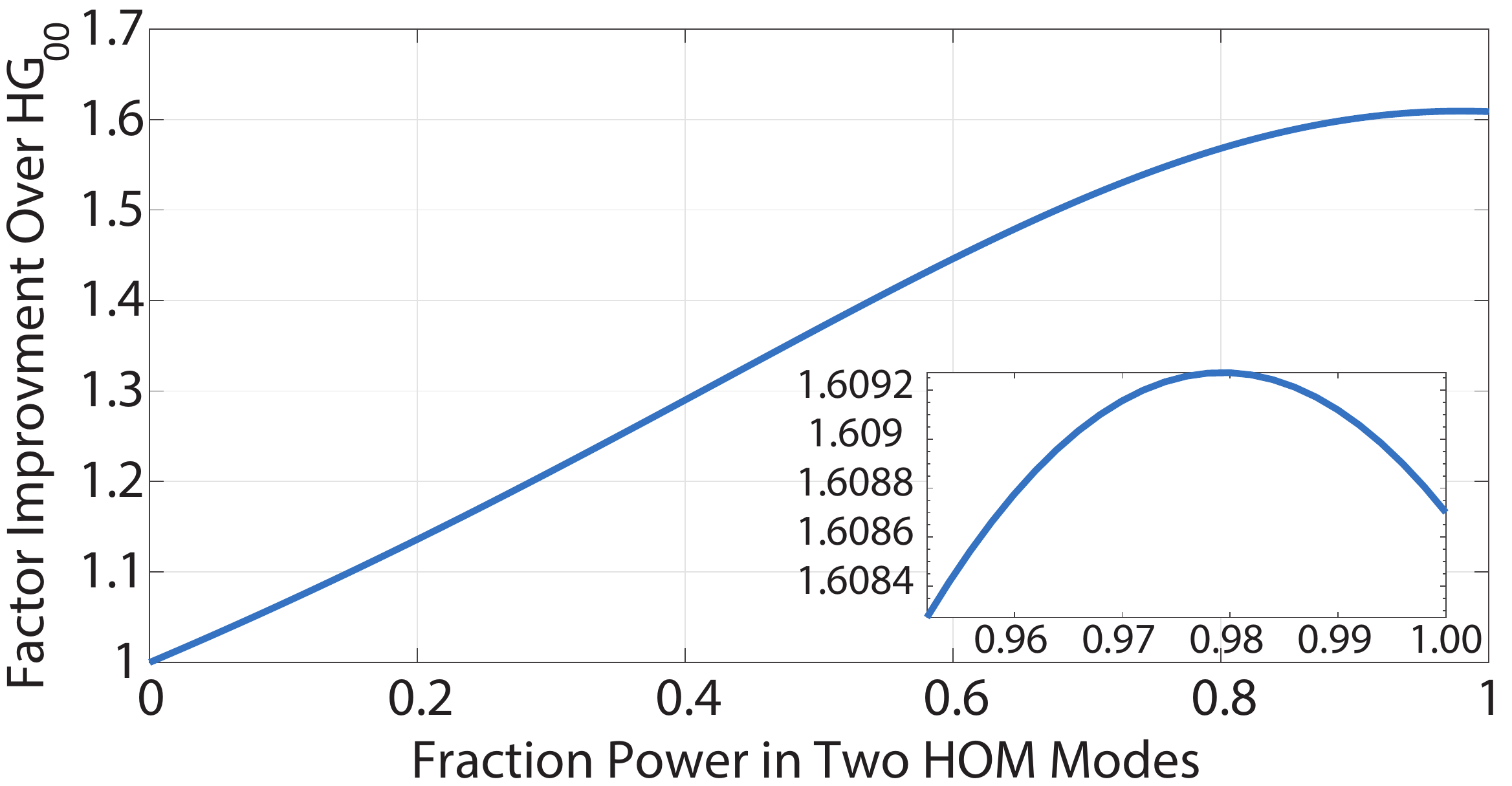}
    \caption{Ratio of thermal noise improvement for a three mode configuration over the fundamental mode.  The faction of total weighting in the two higher modes ($\textrm{HG}_{02}$ and $\textrm{HG}_{20}$) is swept with the remainder in the fundamental mode ($\textrm{HG}_{00}$). Thermal noise improvement peaks at 1.61 with 49\% of the power in each HOM.}
    \label{fig:SweepRatioHOMtoFund}
\end{figure}

Given the symmetry of the mode choices the weighting of the $HG_{02}$ and $HG_{20}$ modes were fixed to be equal with the remainder of the signal weighting given to the fundamental $HG_{00}$.  Thermal noise was computed from a common noise tensor as the weighting of the higher order modes was swept from 0.0 to 1.0.  Here the cavity was 0.1 m long, operating at 1064 nm, with 0.5 m radius of curvature end mirror. The resulting improvement, over the fundamental mode thermal noise, is plotted in Figure \ref{fig:SweepRatioHOMtoFund}. This result shows that for a 0.49 weighting on each of the higher order modes a $1.61\pm0.01$ factor improvement is achievable for a typical fixed spacer reference cavity.  This result is in line with estimates by Zeng et al. (\cite{zeng_thermal-noise-limited_2018}) of 18\% reduction for a single $HG_{02}$ mode.  Our estimate of 38\% reduction is close to double the improvement, a result of the very small overlap of the two modes in the outer lobes of their intensity profiles.\\

In the above result there is a very small (2.0\%) fundamental ($HG_{00}$) component, a result of the strong overlap of the middle peak of the chosen higher order modes with this fundamental mode -- indicating that there is little additional information sampled by the fundamental mode that is not already well sampled by the other two modes. A feedback control scheme using synthesized combined error signals will likely rely entirely on an equally weighted pair of higher order modes.  However, for practical experimental reasons, retaining the $HG_{00}$ mode would likely assist with lock acquisition and diagnostics. Such a three mode scheme could be implemented with off-the-shelf optical components with coupling to the HOM achieved with mode matching of input fundamental modes. Design considerations would need to accommodate for a small allowance for mirror astigmatism to sufficiently separate the frequencies of higher order spatial modes or, alternatively, implement a traveling wave design.  Provided that a suitable separation of higher order mode frequencies can be found, the addition of the $HG_{02} + HG_{20}$ mode pair offers close to additive improvement upon thermal noise reduction compared to locking the cavity with a single $HG_{02}$ mode.\\

\section{Discussion}
Simulations of thermal noise show expected improvements in all  three cases. In Case 1, unlimited injection resource and output signal processing, the improvement appears to be unbounded for short cavities: this reflects the omission of beam clipping that would eventually occur for extremely high mode numbers. While unrealistic, these results show the idealized trend of improvement and highlight the achievable flattening from a superposition of mode readouts.

In Case 2, long aspect ratio cavities, the peak thermal noise improvement, factor 1.57$\pm 0.01$, converges to that of the ideal MESA beam. This also used an unrealistic number of (25) modes. However, a significant improvement of $1.49\pm0.01$ could be seen choosing only even modes (9 in total) of the above 25 modes and warrants further investigation.  In this case we have chosen a 12 m cavity to match the only functional demonstration of MESA beams in optical cavities to date shown in \cite{tarallo_generation_2007}, which was far from demonstrating performance improvements due to manufacturing limitations in the polishing and coating of arbitrarily shaped mirror surfaces\cite{MillerPhD2010}. Of relevance for long baseline gravitational wave detector interferometers (such as Adv. LIGO) is how such a multi-mode technique could be applied for their 4 km arm cavities. Tarallo et al. provide a scaling of MESA beam radius (eq. 4 \cite{tarallo_generation_2007}),
$L_\textrm{prototype} = \left(r_\textrm{mesa}^\textrm{prototype}/ r_\textrm{mesa}^\textrm{AdLIGO}\right)L_\textrm{AdLIGO}$,
with consideration to the clipping limited $HG_{00}$ beam for a 4 km cavity, the MESA improvement achieved is approximately a factor of 2.4 when scaled for 4 km gravitational wave detector cavities \cite{vinet_mirror_2005}. Other beam spreading strategies include the use of higher order Laguerre-Gauss modes, $\textrm{LG}_{33}$ have predict thermal noise reduction by up to 1.7 \cite{chelkowski_prospects_2009} in 4 km cavities, though they are highly susceptible to mirror astigmatism and mode mismatch \cite{sorazu_experimental_2013}.  By contrast higher order Hermite-Gauss modes, $HG_{33}$ modes are expected to have a 1.44 factor of improvement and offer orders of magnitude more robustness than Laguerre-Gauss type modes \cite{tao_higher-order_2020}. Though individual single Hermite-Gauss modes offer less potential improvement, their robustness to manufacturing defects and limitations and the potential for multi-mode combination offers one possible path to lowered thermal noise in addition to existing programs to reduce their coating mechanical losses and cryogenic upgrades.

In short reference cavities the promise of thermal noise reduction through modification of optical mode profiles has only been shown experimentally to reduced thermal noise by 1.2 to 1.4 for $HG_{02}$ and $HG_{24}$ Hermite-Gauss modes \cite{zeng_thermal-noise-limited_2018,NotcuttM2006}. These HOM thermal noise measurements indicate the experimental feasibility of this proposal. The factor 1.6 improvement is equivalent to cooling the system to 120 K from room temperature.  Prospects for non-Hermite Gauss modes for short cavities are limited, given the poor length scaling for robustness to defects in Laguerre-Gauss and other non-Gaussian flattened profile modes (such as MESA). The compound mode approach offers a way to address thermal noise using existing state of the art spherical mirrors.  Challenges in optical engineering are exchanged for injections optics and digital signal processing: areas of development that promise readily available scaling. Future work may also down select mode basis chosen to be less susceptible to these practical experimental considerations while offering a path to thermal noise improvement beyond and in addition to material science advancements and cryogenic upgrades. Future advances in photonic devices for generating multi-higher order mode combinations may make this a viable strategy for long aspect ratio type resonators in the future.

\section{Conclusion}
We proposed an novel strategy for spreading the effective sampling area of beams in optical cavities to reduce coating thermal noise. We analyzed three different scenarios: without beam clipping, a factor 2.46 improvement over $\textrm{TEM}_{00}$ in short cavities for 25 modes; with beam clipping in long 12 m cavities, a factor 1.57 over $\textrm{TEM}_{00}$; and, in an experimentally achievable three mode short reference cavity setup showing a 1.61 factor improvement over $\textrm{TEM}_{00}$.  Such an  approach  utilises existing spherical mirror geometries and is robust against mirror manufacturing errors. Like all thermal noise readout techniques, the improvement offered is independent and in addition to other techniques like low-temperatures and material quality factor improvements.

\begin{acknowledgments}
We wish to acknowledge discussions and advice from Matthew Evans (MIT) and discussion with Ben Buchler and Simon Haine. This research was conducted with support from the Australian Research Council Centre of Excellence for Gravitational Wave Discovery (OzGrav), project number CE170100004, and Centre for Engineered Quantum Systems, project number CE170100009.
\end{acknowledgments}





\bibliography{bibitems}

\end{document}